\documentstyle[prl,aps,twocolumn]{revtex}
\begin{document}
\title{Comment on "Observation of Spin Injection at a Ferromagnet-Semiconductor Interface", by P.R. Hammar et al.}
\author{B.J. van Wees}
\address{Department of Applied Physics and Materials Science Centre, University of Groningen, 9747 AG Groningen, The Netherlands}
\maketitle

In a recent Letter Hammar et al.\cite{ham1} claim the observation of injection of a spin-polarized current in a two-dimensional electron gas (2DEG). This is an important observation, since, despite considerable effort of several groups, all attempts to realize spin-injection into a 2DEG using purely electrical measurements have failed sofar. However, in my opinion  the claim made in \cite{ham1} is not correct, and the observed behaviour can be explained by a combination of a magneto resistance (Hall) effect with a {\it spin-independent} rectification effect due to the presence of a metal-semiconductor junction. 

The interpretation of the data depends crucially on the theoretical description formulated in \cite{john1}. A 2DEG is considered, connected to a ferromagnetic electrode. The key ingredient is that the electron spin is conserved in the 2DEG, but the spin-orbit interaction of the Rashba type induces an asymmetry between electrons moving in a particular direction with different spin directions. As a result rectification is predicted, which depends on the direction $M$ of the magnetization of the ferromagnetic electrode.

However, at low currents linear response theory requires that $V(-I)=-V(I)$. Rectification can only occur for currents $I$ beyond the linear transport regime, under the condition that the transport properties of the electrons are energy dependent, and that the current in both directions is carried by electrons with different energies. However, no energy scale which determines the onset of rectification is discussed in \cite{john1}, where the rectification depends on the direction of the current only. Therefore the theory of \cite{john1} cannot be correct, and it is not possible to detect spin injection in this way.

Turning attention to the experiment, it should be noted first that the authors test the prediction of spin-dependent current rectification by performing a four-terminal measurement. They reverse the current direction by interchanging one current and one voltage lead. This however is not a critical test of the theory, since \cite{john1} predicts a change in resistance when the current direction is reversed, {\it but the current and voltage leads themselves are not changed}. The reciprocity theorem for multi-terminal measurements \cite{but1} states that (in the linear transport regime) the resistance should be invariant under interchange of the current and voltage leads, accompanied by a reversal of the magnetic field $B$. Although in the experiment only one current and voltage probe is interchanged, the result should be almost equivalent to interchanging both pairs, since the other pair is connected by the low resistance ferromagnetic electrode, and interchanging this pair should make little difference on the measured resistance. Therefore the observed behaviour is consistent with the reciprocity theorem, and the change in resistance can possibly be caused by a Hall effect due to the reversal of the magnetic field (e.g. resulting from the fringe fields at the edges of the ferromagnetic electrode). An estimate of the effect on the resistance is difficult however, due to the uncertainty where the current from the 2DEG actually enters the ferromagnet.

The second  important aspect is that the authors study transport through a semiconductor-metal system, which contains a barrier. Judged from the measured resistance, which is low and comparable to the resistance of the 2DEG itself,  the barrier is not very effective. However, on the voltage scale of $0.1 V$ which the authors use, some sort of rectification might occur, leading to $V(-I)\neq -V(I)$. This effect will {\it not} depend on the direction of the magnetization {\it M}, since the electron spin is not relevant for rectification in metal-semiconductor junctions. In other words: $V(I)_{M}=V(I)_{-M}$. 

The presented data therefore does not exclude that the observed effects are due to a combination of a (magnetization independent) rectification, combined with a magneto resistance (Hall) effect. It is however possible to distinguish between this explanation and that of the authors.  To support their case they should show that the rectification behaviour observed in the $V(I)$ characteristics changes sign, when the magnetization $M$ is reversed.  In other words, for a fixed configuration of current and voltage leads, they should show explicitly that the relation  $V(I)_{M}= -V(-I)_{-M}$  is obeyed, for the full current range. In particular this implies that when $M$ switches direction when the coercive field is exceeded, and a change $\Delta R$ is observed for positive $I$, an {\it opposite} change $-\Delta R$ should be observed at negative $I$. This behaviour can then clearly be distinguished from a Hall effect, which predicts that the  change in resistance $\Delta R$ will have the same sign for both current directions.

\end{document}